# Bending of Sparklers

Mikrajuddin Abdullah[1,a], Shafira Khairunnisa[2], and Fathan Akbar[3]

[1]*Department of Physics, Bandung Institute of Technology,*

*Jl. Ganesha 10 Bandung 40132, Indonesi*a

[2]*Department of Food Technology, Padjajaran University, Jalan Raya Jatinagor, Sumedang, Indonesia*

[3]*SMAN 24 Bandung, Jl. A.H. Nasution, Bandung, Indonesia*

[a]Corresponding author: din@fi.itb.ac.id

Abstract

A new equation is proposed to explain the curvature of spent sparklers. We found the state of a segment of the sparkler to depend strongly on the state of its spent segments. The equation is nearly able to produce the sparkler shape for a range of lengths and for all elevation angles. The method proposed here is likely to explain any phenomena in nature related to an evolving length scale associated with some material that becomes progressively stiff or dry, such as the growth of resin exuded from trees. The equation can produce a very rich spectrum of shapes by varying material parameters (density, temperature-dependent strength), heating temperature, elevation angle, and gravitational acceleration. This might provide new insights into explaining many shapes in nature or man-made structures.

# I. INTRODUCTION

The most popular entertainment for children on the night of end-of-year parties or the end of Ramadan is fireworks. A delight for most children is the hand-held sparkler, a stiff metal wire partially coated with slow-burning chemical composition that when lit emits a colourful sparkling spectacle (the uncoated portion is required for its safe handling) [1]. The sparkler is lit from its free end to generate a hot spot that moves down the length of the chemically coated part of the wire. The initially straight sparkler curls during burning. Observing many spent sparklers, we found various curvatures depending on how the sparkler is held during burning. Hence it is interesting to develop a mathematical model to match the transformation in the shape of the sparkler as it burns.

The main metal used to make fireworks for children is aluminium alloy. The temperature of the hot spot is high enough to soften the metal wire. Because of gravitational torque, the weight of the already spent segments bends the softened segment (assume the wire is divided into a number of similar segments), resulting in the wire curling. The bend angle should increase as the spent portion lengthens because of the larger torque. After the softened segment solidifies, the angle of this segment relative to the previously spent segment is fixed. This phenomenon is vastly different from bending a metal beam supported at one end [2]. The angle between the nearest segments in the beam (assuming the beam is divided into a number of similar segments) can vary depending on the mechanical stability.

Although discussion here was focused on sparklers bending, however, the model might be used to explain other phenomena in nature where the evolution of a length of material depends on an asymmetric one-sided transformation in some property of the material.

# II. THEORY

The purpose of this work is to develop a mathematical formulation for the sparkler as it bends and to perform some experiments to confirm the simulated results. The structure of the sparkler reported here is illustrated in Fig. 1 (a). It consists of a metal rod/wire partially coated with a slow-burning chemical composition to a length $L$ along the wire. We divide the coated portion into $N$ equal segments, the length of each then being $s = L/N$.

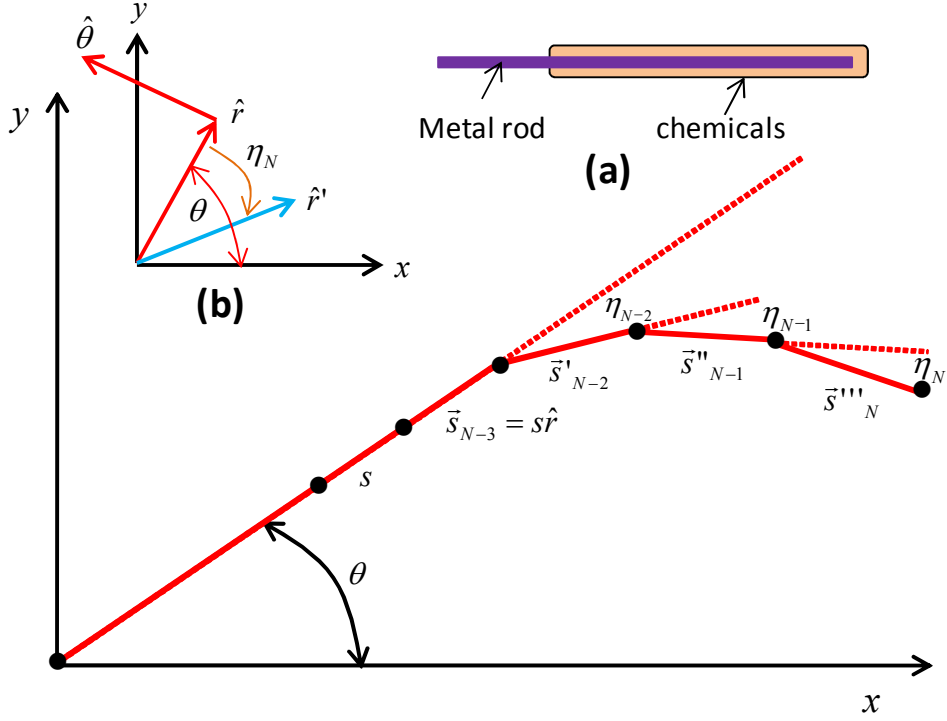

FIG 1. (a) A schematic of a sparkler. The large curve shows the chemical portion of the sparkler divided into $N$ segments of equal length. Each segment is denoted by a vector. During burning, the segments bend sequentially from the free end towards the held end. We assume the time step for burning is proportional to the number of segments. b) Orientation of the $N$-th segment before and after bending. Bending can be considered as a clockwise rotation of the vector.

The sparkler is elevated at an angle $\theta$ to the horizontal; at $t = 0$ (start of burning), all segments are equally oriented. The vector of each segment is $\vec{s}_k = s\hat{r}$ ($k = 1$ (leftmost) to $N$ (rightmost)) with

$$\hat{r} = \hat{i}\cos\theta + \hat{j}\sin\theta. \tag{1}$$

The segments are bent sequentially for each increment in time. At the $t = 1$ time unit, the $N$-th segment bends through angle $\eta_N$, corresponding to a clockwise rotation by $\eta_N$, and the new direction of this vector is

$$\vec{r}\,' = \hat{r}\cos\eta_N - \hat{\theta}\sin\eta_N, \tag{2}$$

with

$$\hat{\theta} = \hat{r}(\theta + \pi/2) = -\hat{i}\sin\theta + \hat{j}\cos\theta. \tag{3}$$

We assume the magnitude of bending angle to be proportional to the torque acting on the spent segments. For this purpose we start by determining the distances of center of mass of the spent segment at different time units:

at $t = 0$,

$$x_c(0) = \frac{1}{2}s\cos\theta,$$

at $t = 1$ time unit

$$x_c(1) = \frac{\Delta m\left[\frac{1}{2}s\cos\theta\right] + \Delta m\left[s\cos\theta + \frac{1}{2}s\cos(\theta - \eta_N)\right]}{2\Delta m}$$

$$= \frac{1}{2} \times \frac{1}{2}s[3\cos\theta + \cos(\theta - \eta_N)],$$

at $t = 2$ time unit

$$x_c(2) = \frac{1}{3\Delta m}\left\{\Delta m\left[\frac{1}{2}s\cos\theta\right] + \Delta m\left[s\cos\theta + \frac{1}{2}s\cos(\theta - \eta_{N-1})\right]\right.$$

$$\left. + \Delta m\left[s\cos\theta + s\cos(\theta - \eta_{N-1}) + \frac{1}{2}s\cos(\theta - \eta_{N-1} - \eta_N)\right]\right\}$$

$$= \frac{1}{3} \times \frac{1}{2}s[5\cos\theta + 3\cos(\theta - \eta_{N-1}) + \cos(\theta - \eta_{N-1} - \eta_N)],$$

etc. Therefore, the magnitudes of bending angles for different segments are

$$\eta_N = \gamma(\Delta mg)x_c(0) = \gamma(\mu sg) \times \frac{1}{2}s\cos\theta = \kappa s^2 \cos\theta \tag{4}$$

$$\eta_{N-1} = \gamma(2\Delta mg)x_c(1) = \kappa s^2 [3\cos\theta + \cos(\theta - \eta_N)] \tag{5}$$

$$\eta_{N-2} = \gamma(3\Delta mg)x_c(2) = \kappa s^2 [5\cos\theta + 3\cos(\theta - \eta_{N-1}) + \cos(\theta - \eta_{N-1} - \eta_N)] \tag{6}$$

etc., with $\Delta m = \mu s$ the mass of each segment, $\mu$ the mass per unit length of burnt part, $g$ the gravitation acceleration, $\gamma$ the constant of proportionality, which depends on the material properties of the metal rod, and $\kappa = \gamma\mu g/2$. Clearly, from Eqs. (4) - (6), the magnitude of bending at a certain segment depends on the magnitude of bending of all preceding segments, or $\eta_k = f(\eta_{k+1}, \eta_{k+2}, ..., \eta_N)$.

After burning has finished, each segment has rotated by different angles; the $k$-th segment has rotated to $\sum_{n=1}^{k}\eta_n$, or made an angle $\theta - \sum_{n=1}^{k}\eta_n$ to the horizontal. To draw the final shape of the spent sparkler, we must determine the positions of all segments. We can easily show that the coordinates of the bent segments are

$$x_k = x_{k-1} + s\cos\left(\theta - \sum_{n=1}^{k}\eta_n\right), \tag{7}$$

$$y_k = y_{k-1} + s\sin\left(\theta - \sum_{n=1}^{k}\eta_n\right), \tag{8}$$

with $x_0 = y_0 = 0$. This is the new iterative equation unreported until now. Equations (7) and (8) express the state of a given segment as being determined by states of all spent segments on the free end. These equations are vastly different from the cellular automata equations, where the state of a point generally depends on the states of its nearest neighbour [3-5].

The experiments were performed by fixing the sparklers at desired angles. Firing was started at the outmost edge using a common kitchen match. The total time a 0.3-m sparkler burns is around 89 s (giving a burning speed of around $3.37 \times 10^{-3}$ m/s). The completely spent sparklers were then pictured using a digital camera and made clear using an image processing software.

## III. RESULTS AND DISCUSSION

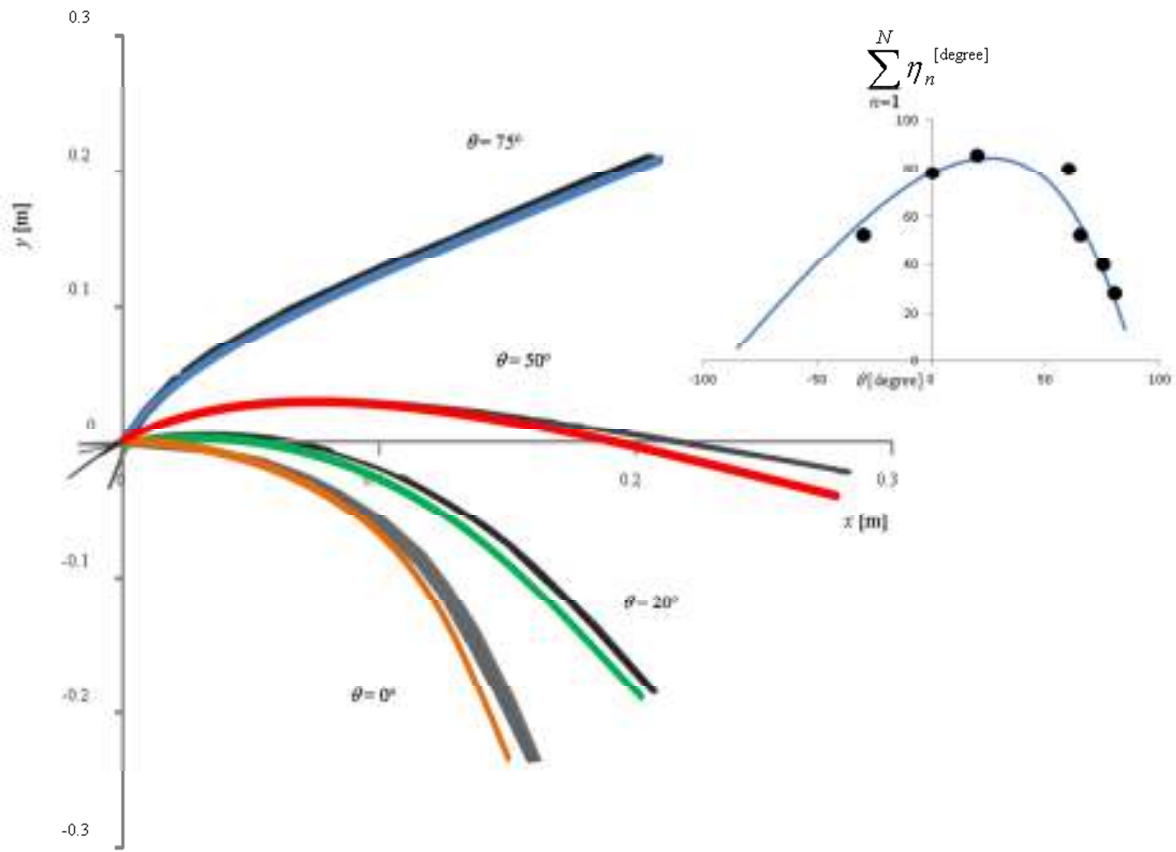

FIG. 2. Comparison of observed (black curves) and simulated (color curves) shapes of spent sparklers at different elevation angles ($\theta$): 0°, 20°, 50°, and 70°. All curves were calculated using $\kappa = 0.131$ rad/m². (inset) Comparisons of the simulated (line) and measured (symbols) total bending angles of the $N$-th segment.

In this work we used a sparkler having chemically coated part of its 30-cm length and an aluminium alloy rod of diameter 1.5 mm. Figure 2 shows a comparison between experimental observations of the bending of spent sparklers with simulated results of sparklers elevated at different angles. In simulations, the coated portion of the sparkler was divided into 400 segments (the length of each was 0.00075 m). We selected a $\kappa$ value so that the experimental and simulated shapes for a certain elevation angle well matched each other. The chosen $\kappa$ was then used to compute all simulated curves. The best value for this parameter was $\kappa = 0.131$ rad/m$^2$, because it was able to match the observed bending of sparklers that have been positioned at different elevation angles and sparklers of different lengths. The combination of $\kappa$ and $s$ values are very important. Different values might well match the bending of 0.3 m length sparkles having different elevation angles, but fail to match the sparklers of different length, and vice versa. All curves showed good agreement with the experimental data, especially for large angles of elevation.

After burning was completed, all segments developed different total bending angles relative to their initial directions. Inset in Fig. 2 presents a comparison between the measured and simulated total bending angles of the $N$-th segment. For small and large elevation angles, good agreements were obtained, although for medium elevation angles (around 0° to 60°) the simulated results are slightly lower than the observed results. However, both observation and simulated data showed similar trends, having a maximum bending of around $\theta = 27°$. This result confirmed the proposed model is useful for predicting bending in sparklers.

The final shape of spent sparklers should depend on their length. However, it is difficult to obtain very long sparklers in the market; the longest are of the order of tens of centimetres. Indeed, it is challenging to reproduce the shape of very long sparklers. To model such sparklers, we might use a supported metal rod of the same material placed in such a way that the elevation angle is fixed at all points (a freely stuck long rod easily bends downward because of gravity). The supported rod is then heated by the hot spot moving progressively away from the free end. The temperature of the heat source should be equal to the temperature of the hot spot of the sparkler and the speed of the downward motion of the heat source should be equal to the speed the hot spot moves along the sparkler.

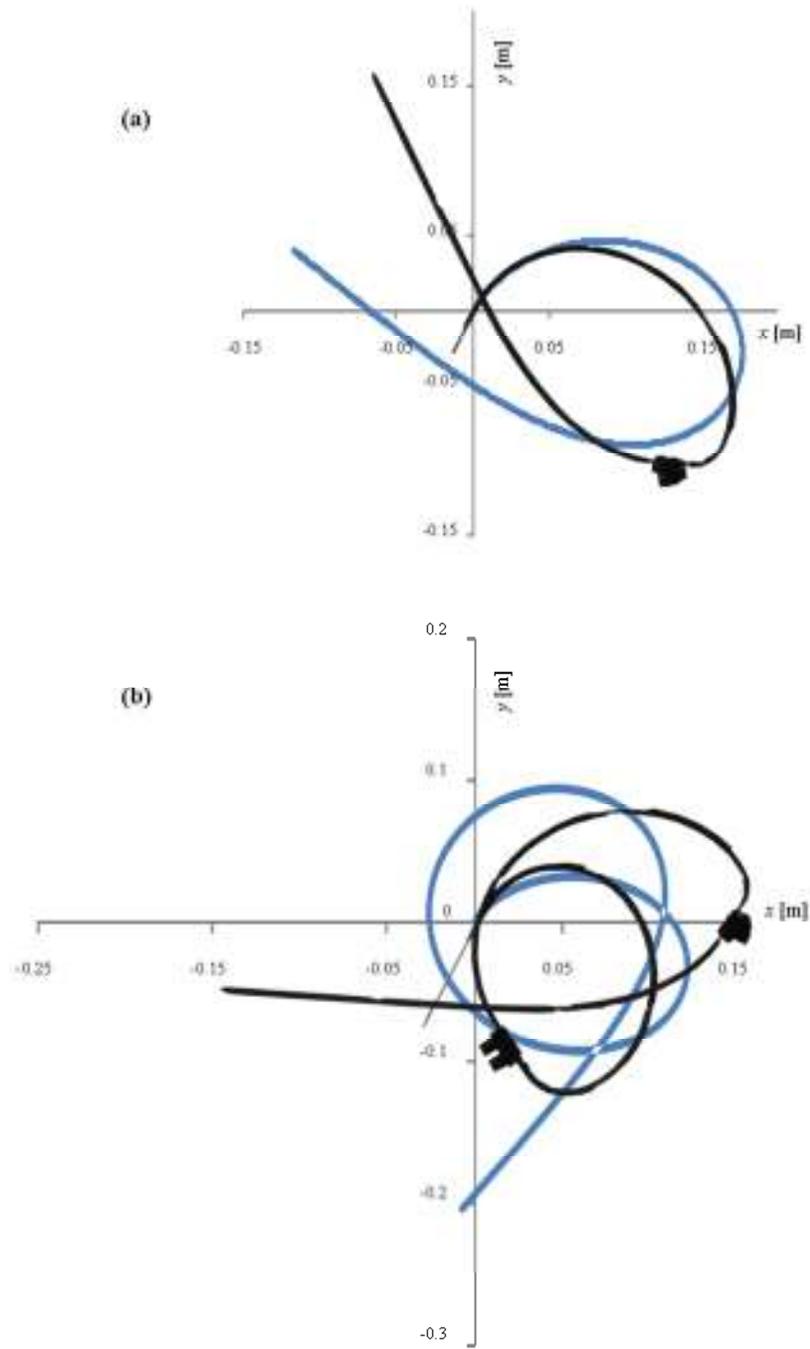

FIG 3. Simulated (color curves) and observed (black curves) of final shapes of spent sparklers of different lengths elevated at $\theta = 50°$: (a) $L = 0.6$ m and (b) $L = 0.9$ m. These observed sparklers were obtained by joining two and three smaller sparklers.

Figure 3 shows the simulated shapes of fireworks of different lengths fixed elevated to $\theta$ = 50° (color curves). The length of each segment is fixed at 0.00075 m. In a rough demonstration, we proved the presence of this shape by joining two and three fireworks using an electrical joiner, usually used to join conductor cable. Indeed, the presence of the joiner may have altered the homogeneity of chemical portion of the sparkler; hence we were unable precisely to reproduce the observed shapes. However, we observed qualitatively a similar shape transformation in the spent sparklers (black curves).

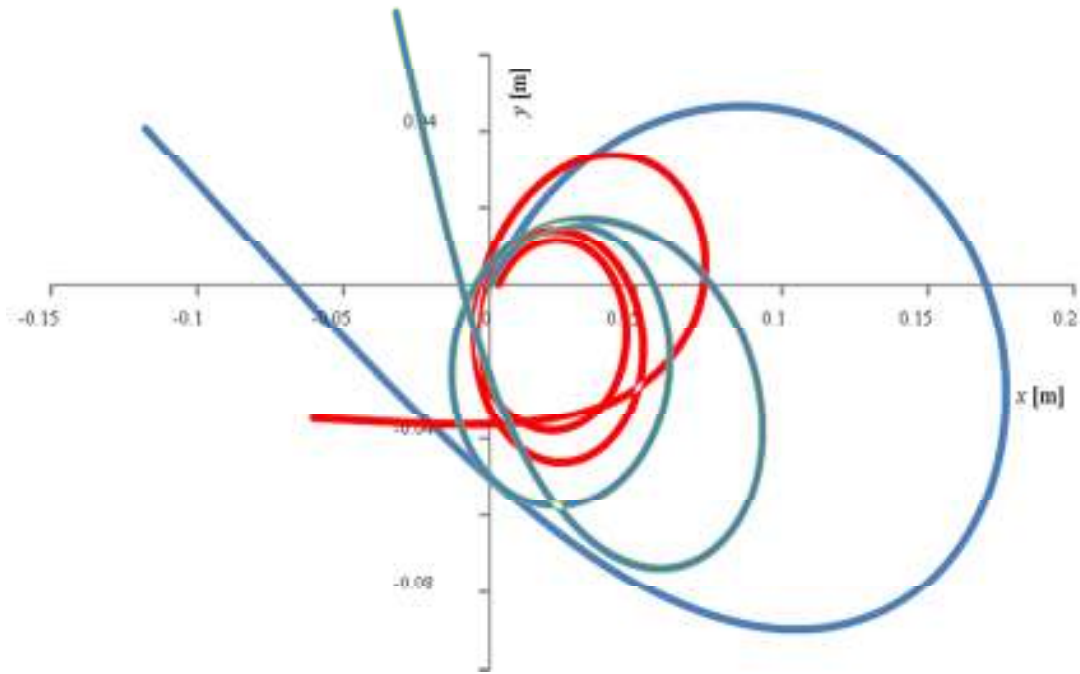

FIG. 4. Simulated shapes of sparklers of length $L$ = 0.6 m, and elevated by $\theta$ = 60°, for different $\kappa$: (blue) 0.16 rad/m², (green) 0.8 rad/m², and (red) 1.6 rad/m². Higher $\kappa$ values correspond to either higher hot spot temperatures or higher "gravitational" accelerations, and vice versa.

The next query is what happens when the temperature of the hot spot is increased or lowered. When the temperature of the hot spot is increased, the hot spot of the rod becomes softer, which induces a greater bending angle under the same applied torque. This situation corresponds to a larger $\kappa$ constant. Conversely, when the temperature is lower, the metal rod becomes stiffer and the same torque acting on the free end results in smaller bending angles. Because $\kappa = 0.5\gamma\mu g$, similar shapes will also be observed under different gravitational accelerations. Smaller g-values correspond to stiffer hot spots (i.e., lower heating temperatures).

Figure 4 presents the shape of spent sparklers of 0.6 m length with different $\kappa$ (either different temperatures at the hot spot or different gravitational accelerations). The effective curling radius decreases when κ increases. Very large $\kappa$ corresponds to a very soft hot spot, causing large bending to occur even for small torques.

## IV. RESUME

The model proposed here can explain other phenomena in nature where the evolution of a length of material depends on an asymmetric one-sided transformation in some property of the material. An example of this phenomenon is the growth of resin exuded from a tree. The viscous resin secreted from certain trees is soft and hardens over a certain time period. The length of the hardened resin increases as more resin is exuded, pushing out the hardened resin. Gravitational forces acting on the hardened resin cause the resin to bend. The final shape of long hardened resin is nearly similar to shapes simulated here.

Finally, Eqs. (7) and (8) can produce very rich shapes by varying the material parameters (density, temperature-dependent strength), heating temperature, elevation angle, and gravitational acceleration. This might provide new insights into explaining many shapes in nature or man-made structures.

**Supported materials can be obtained from:**

Video when elevated at 60 degree

http://www.mediafire.com/watch/0v1ruq2czrluuwu/Sparklers%20elevated%20at%2060%20degree.flv

Video at zero degree elevation

http://www.mediafire.com/watch/up9hxydyut4x4mn/Sparklers%20of%20zero%20elevation.flv